\documentclass[iop, revtex4]{emulateapj}

\usepackage{graphicx}
\usepackage{color}
\usepackage{amsmath}
\usepackage{natbib}
\usepackage{float}
\usepackage[caption = false]{subfig}
\usepackage[utf8]{inputenc}
\usepackage[T1]{fontenc}
\usepackage{fancyhdr}

\newcommand{\degree}{$^{\circ}$}

\newcommand{\be}[1]{\begin{equation} \label{eq:#1}}
\newcommand{\ee}{\end{equation}}

\newcommand{\solrad}{\ifmmode{R}_{\rm S}\else${R}_{\rm S}$\fi}
\newcommand{\solmas}{\ifmmode{M}_{\rm S}\else${M}_{\rm S}$\fi}

\newcommand{\tintu}{\ifmmode{\rm erg~cm^{-2}~s^{-1}sr^{-1}}\else 
 erg~cm$^{-2}$~s$^{-1}$~sr$^{-1}$\fi}
\newcommand{\fluxu}{\ifmmode{\rm erg~cm^{-2}~s^{-1}}\else 
  erg~cm$^{-2}$~s$^{-1}$\fi}

\newcommand{\wave}{\ifmmode{\lambda} \else$\lambda$\fi}

\newcommand\lta { \mathrel {\hbox to 0pt {\lower 3.7pt \hbox{$\sim$}
      \hss} \raise 1.7pt \hbox{$<$}}}
\newcommand\gta { \mathrel {\hbox to 0pt {\lower 3.7pt \hbox{$\sim$}
      \hss} \raise 1.7pt \hbox{$>$}}}
\newcommand{\lasco}{The LASCO CME catalog is generated and maintained at the CDAW Data Center by NASA and The Catholic University of America in cooperation with the Naval Research Laboratory. SOHO is a project of international cooperation between ESA and NASA.}

\begin{document}

%
%

\title{Multiwavelength Stereoscopic Observation of the May 1, 2013 Solar Flare and CME}
\submitted{\footnotesize{Accepted by The Astrophysical Journal on 2019 September 30 \\ https://doi.org/10.3847/1538-4357/ab4a0a }}

\author{Erica Lastufka}
 \affil{Fachhochshule Nordwestschweiz,\\
      Bahnhofstrasse 6, 5210 Windisch, Switzerland;}
\affil{ETH-Zürich,\\
      Rämistrasse 101, 8092 Zürich, Switzerland;}

\author{S\"am Krucker}
 \affil{Fachhochshule Nordwestschweiz,\\
      Bahnhofstrasse 6, 5210 Windisch, Switzerland;}
\affil{Space Science Laboratory, UC Berkeley\\
      Berkeley, USA;}

\author{Ivan Zimovets}
\affil{Russian Academy of Sciences (IKI),\\ 84/32 Profsoyuznaya Str, Moscow, Russia, 117997} 

\author{Bulat Nizamov}
\affil{Lomonosov Moscow State University,\\ Universitetskii Pr. 13, Moscow, Russia, 119992} 

 \author{Stephen White}
  \affil{Air Force Research Laboratory, \\ Albuquerque, NM, USA}

 \author{Satoshi Masuda}
  \affil{Nagoya University, \\ Nagoya 464-8601, Japan}

\author{Dmitriy Golovin}
 \affil{Russian Academy of Sciences (IKI),\\ 84/32 Profsoyuznaya Str, Moscow, Russia, 117997} 
\author{Maxim Litvak}
 \affil{Russian Academy of Sciences (IKI),\\ 84/32 Profsoyuznaya Str, Moscow, Russia, 117997} 
\author{Igor Mitrofanov}
 \affil{Russian Academy of Sciences (IKI),\\ 84/32 Profsoyuznaya Str, Moscow, Russia, 117997} 
\author{Anton Sanin}
 \affil{Russian Academy of Sciences (IKI),\\ 84/32 Profsoyuznaya Str, Moscow, Russia, 117997} 

\begin{abstract}

A M-class behind-the-limb solar flare on 1 May 2013 (SOL2013-05-01T02:32), accompanied by a ($\sim$ 400 km/s) CME was observed by several space-based observatories with different viewing angles.
We investigated the RHESSI-observed occulted hard X-ray emissions that originated at least 0.1 \solrad{} above the flare site. Emissions below $\sim$10 keV revealed a hot, extended (11 MK, >60 arcsec) thermal source from the escaping CME core, with densities around $10^{9}$ cm$^{-3}$. In such a tenuous hot plasma, ionization times scales are several minutes, consistent with the non-detection of the hot CME core in SDO/AIA's 131 \AA{} filter. 
The non-thermal RHESSI source originated from an even larger area ($\sim$100 arcsec) at lower densities ($10^{8}$ cm$^{-3}$) located above the hot core, but still behind the CME front. This indicates that the observed part of the non-thermal electrons are not responsible for heating the CME core. Possibly the hot core was heated by non-thermal electrons before it became visible from Earth, meaning that the un-occulted part of the non-thermal emission likely originates from a more tenuous part of the CME core, where non-thermal electrons survive long enough to became visible from Earth.
Simultaneous hard X-ray spectra from the Mars Odyssey mission, which viewed the flare on disk, indicated that the number of non-thermal electrons $>$20 keV within the high coronal source is $\sim$0.1 - 0.5\% compared to the number within the chromospheric flare ribbons. 
The detection of high coronal hard X-ray sources in this moderate size event suggests that such sources are likely a common feature within solar eruptive events.

\end{abstract}

\section{Introduction}

Coronal hard X-ray (HXR) sources provide one of the most exciting diagnostics of coronal plasma. However, the majority of flare-associated HXRs that we can observe occur at the footpoints of flare loops, when accelerated electrons collide with dense chromospheric plasma. 
Because these sources are orders of magnitude brighter than coronal HXR emission, most X-ray instruments are incapable of resolving both coronal and chromospheric sources simultaneously. 


Coronal X-ray emission can have both thermal and non-thermal components. These can be difficult to distinguish from each other in the low corona, where accelerated electrons interact with dense arcades of cooling flare loops. 
For less powerful or more compact flares, looptop sources, or indeed any X-ray source located even higher up in the corona (hard X-ray sources have been observed up to 0.3\(R_\odot\)), are only observable when the flare footpoints are occulted behind the solar disk.
Only in certain spectacular cases, such as the Masuda flare \citep{masudaLooptopHardXray1994}, can the thermal and non-thermal components be well-separated spatially by dynamic-range limited intsruments, without the additional advantage of footpoint occultation. 

Observations of both occulted flares and Masuda-type events have had profound impacts on our understanding of particle acceleration in flares.
 Mechanisms of particle acceleration, trapping, or turbulence that are responsible for producing bremsstrahlung in what is traditionally thought of as tenuous coronal plasma are not yet well understood. Multi-wavelength observations of coronal emission, as well as of the complete flare, are the keys to characterizing particle acceleration from one region to the next.

Unfortunately, comprehensive observations of occulted flares, especially those with X-ray sources very high in the corona, are rare. Two of the best-observed examples are the X-class flare of October 27, 2002 \citep{kruckerSolarFlareHard2007} and the smaller\footnote{The unocculted SXR flux registered as GOES C5 class. Chertok's method gave a class of X1.5.} flare of November 3, 2010 \citep{glesenerObservationHeatingFlareaccelerated2013a}. The powerful October 27 flare showed an extended coronal source that moved rapidly (750 km/s) in the same direction as the accompanying coronal mass ejection (CME). An estimated 10\% of the electrons in the source were determined to be nonthermal, possibly particles trapped on field lines related to the CME. 

Another facet of one global magnetic eruption, CMEs, which often accompany flares \citep{harrisonNatureSolarFlares1995,zhangTemporalRelationshipCoronal2001}, can contribute to the presence of locally dense plasma capable of producing enhanced bremsstrahlung.
As a flux rope rises from the solar surface, a current sheet forms below it, containing microscopic instabilities resulting in fast magnetic reconnection. 
Following reconnection, the envelope of field lines overlying the flux rope is pushed up, forming the CME's bright frontal loop with a piston-driven shock ahead of it. Part of the flux rope is now known as the CME core, while the remainder falls to the solar surface (Gopalswamy et al., 2003; Jing et al., 2004). High-energy particles and hot plasma can be released in bi-directional outflows observable in radio and X-rays \citep{liuDoubleCoronalHard2008}.
Their motion can often be traced by co-spatial soft X-ray (SXR) or EUV plasmoids that form as the upward reconnection flow hits the CME core. 

\begin{figure}
\includegraphics[width=0.49\textwidth]{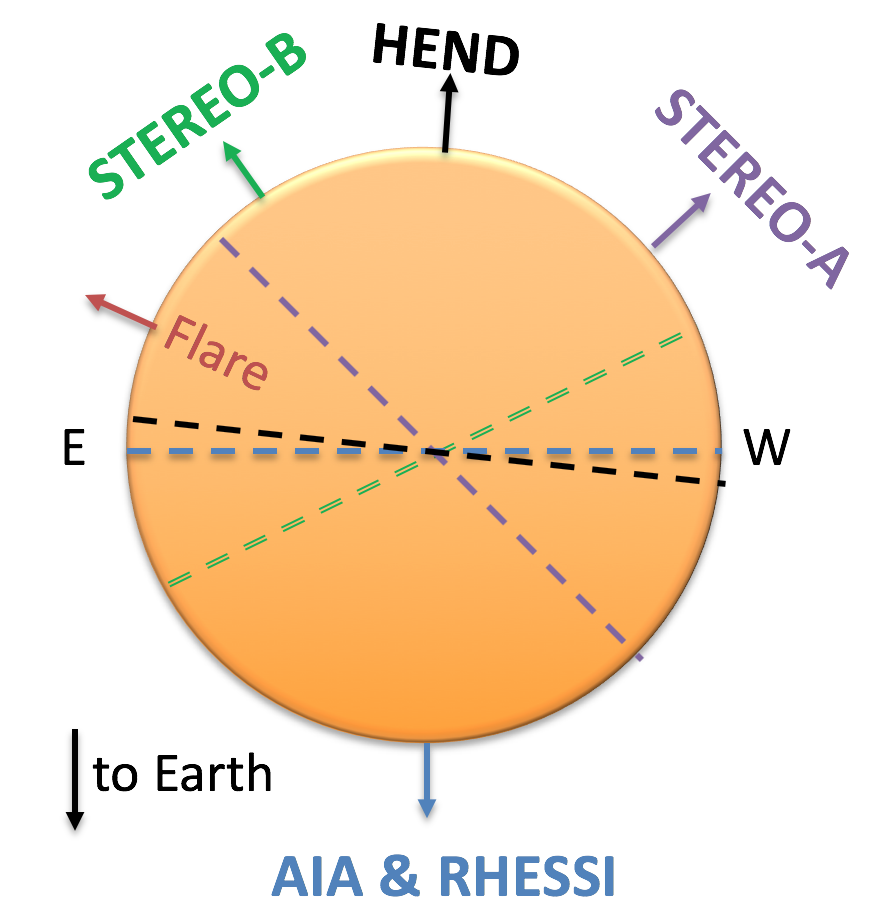}
\caption{The flare of May 1, 2013, was observed from many perspectives. This diagram imagines looking down on the ecliptic; solid vectors point towards the satellites, and dashed lines represent the half of the Sun visible from each perspective.} 
\label{fig:geom}
\end{figure}

\begin{figure*}
\includegraphics[width=\textwidth]{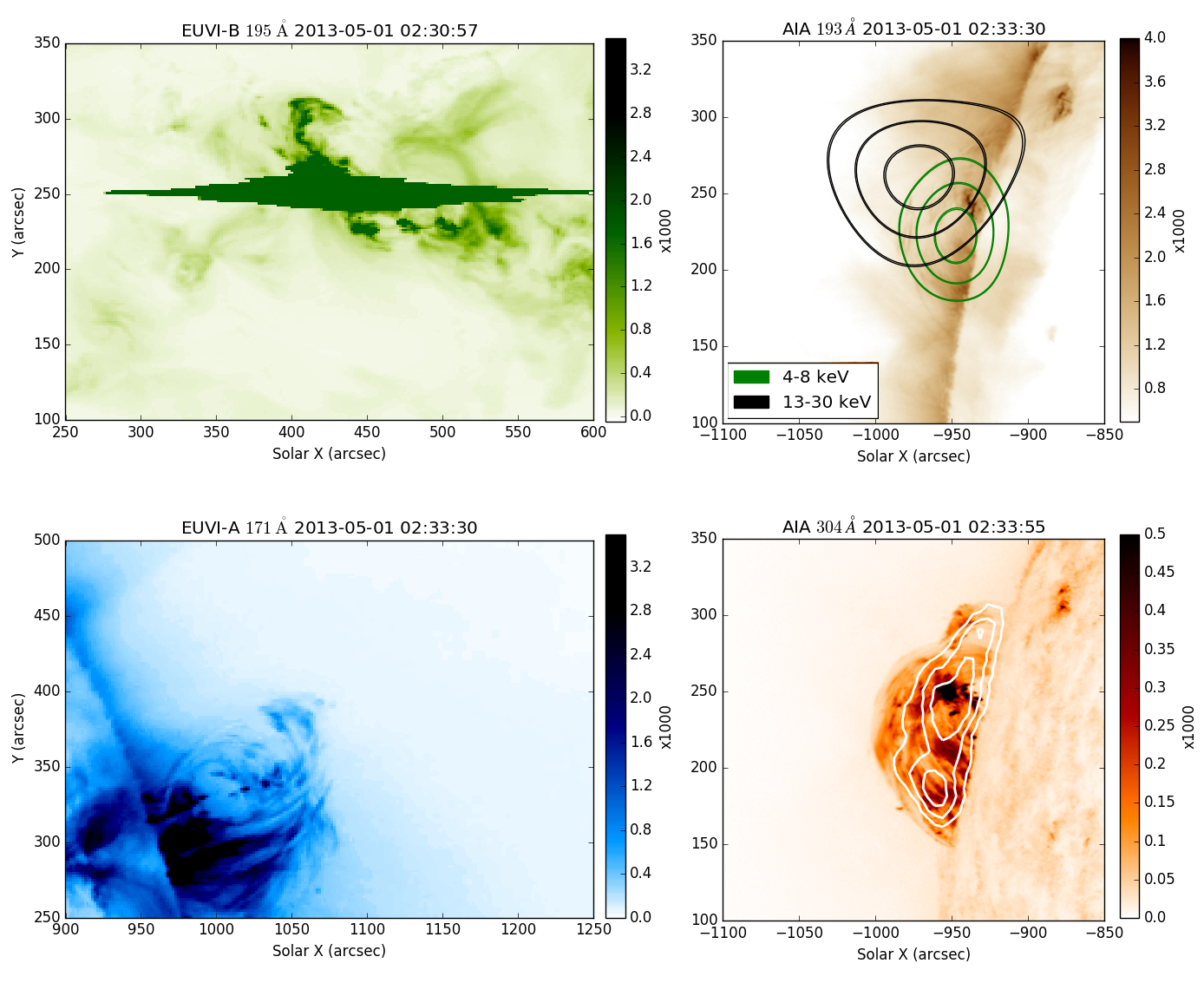}
\caption{From top-left, clockwise: The flare peak observed directly by STEREO-B 195 \AA. RHESSI imaging of the flare peak at 02:33UT showed an extended non-thermal source leading a compact thermal source. No corresponding sources were visible in AIA 193 \AA. However, plasma observed at 17 GHz by NoRH resembled the cooler-temperature AIA 304 \AA{} images. The view from STEREO-A was occulted by 3\degree{} less than AIA, so it saw slightly more of the flare. An animated version of this figure is available,  following the flare evolution. From an hour before the flare erupts, the full-disk view offered by STEREO-B shows the active region changing. During and after the flare peak, short-lived X-ray emission and longer-lived radio emission is seen. The filament eruption is seen especially well from the STEREO-A viewpoint, with EUV plasmoids stretched along the filament axis seven minutes after the peak.
} 
\label{fig:movie}
\end{figure*}

Hard X-rays have also been observed moving with CME cores \citep{hudsonHardXRadiationFast2001}. 
Flare-accelerated electrons in the CME core can cause the thermal CME energy to exceed the CME kinetic energy \citep{leeThreeDimensionalStructureEnergy2009,landiPhysicalConditionsCoronal2010}, resulting in nonthermal electron energy loss, observed through HXR emission. 
The November 3 flare studied by \citet{glesenerObservationHeatingFlareaccelerated2013a} is a prime example - simultaneous EUV and X-ray observations showed that flare-accelerated electrons had enough energy to heat the CME core.
It is clear that the dynamics of CMEs and the energy release of flares are intrinsically coupled to each other. 

The event of May 1, 2013 (SOL2013-05-01T02:32) is an excellent opportunity to investigate how and where a solar eruption accelerates particles. Observed by an array of instruments ranging in wavelength from radio to hard X-ray, it enables the study of the coronal electron population at many temperatures.
In this paper, we focus on both the potential causes of the observed X-ray emission and its properties.

\section{Event Description}
\subsection{Observation Geometry}

The active region that produced the May 1, 2013 flare was fully visible to STEREO-B's SECCHI \citep{howardSunEarthConnection2008}, to the north-east of its field of view. The centroid of the peak flare image gave a precise location of (424",241"), which allowed us to infer the observation geometry. The flare position was located 30\degree{} behind the east solar limb as seen from Earth, which meant that the observed plasma was emitted from heights at least 111 Mm radially above the active region. Mars Odyssey saw almost the opposite view from that of AIA, with the flare seen close to the limb but on disk. 
STEREO-A saw the flare as occulted by 27\degree{} behind STEREO-A's east limb, which allowed plasma 88 Mm or more above the flaring region to be observed. 
Figure \ref{fig:geom} shows the approximate location of the flare with respect to the various observatories, and their fields-of-view at the time. 

\subsection{Event Overview}\label{sec:description}

\begin{figure*}[ht]
\includegraphics[width=.64\textwidth]{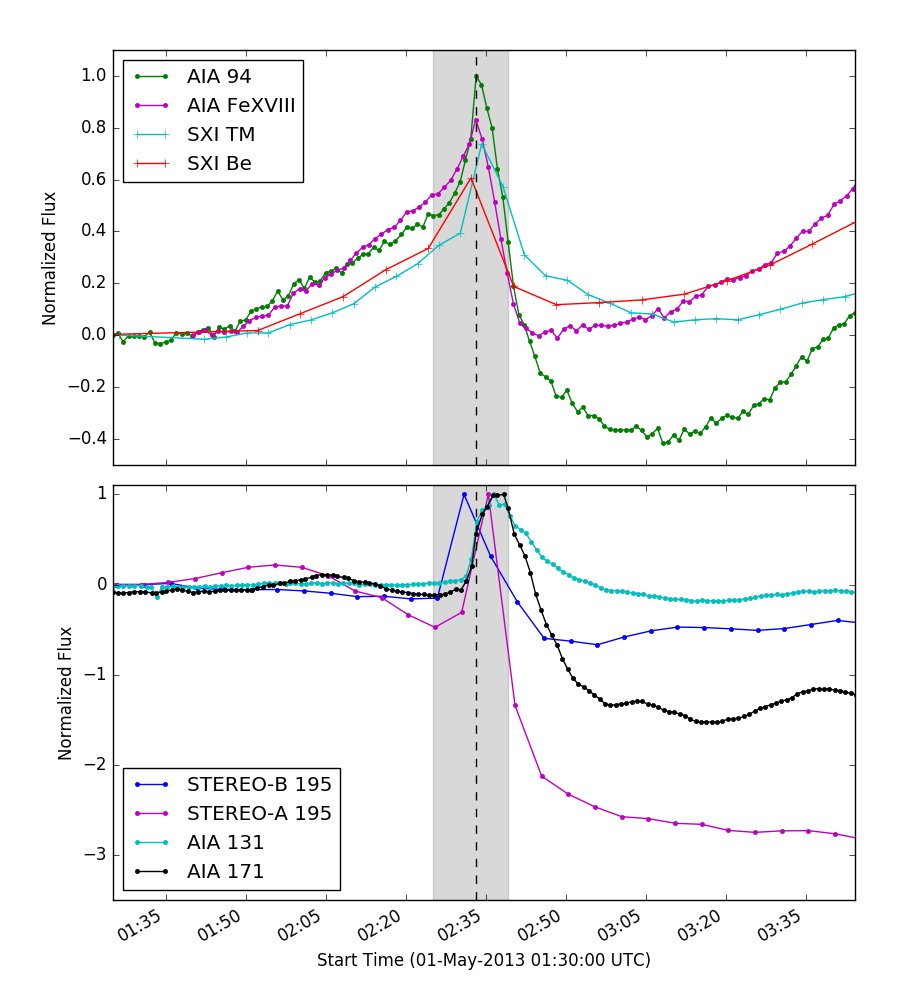}
\includegraphics[width=.36\textwidth]{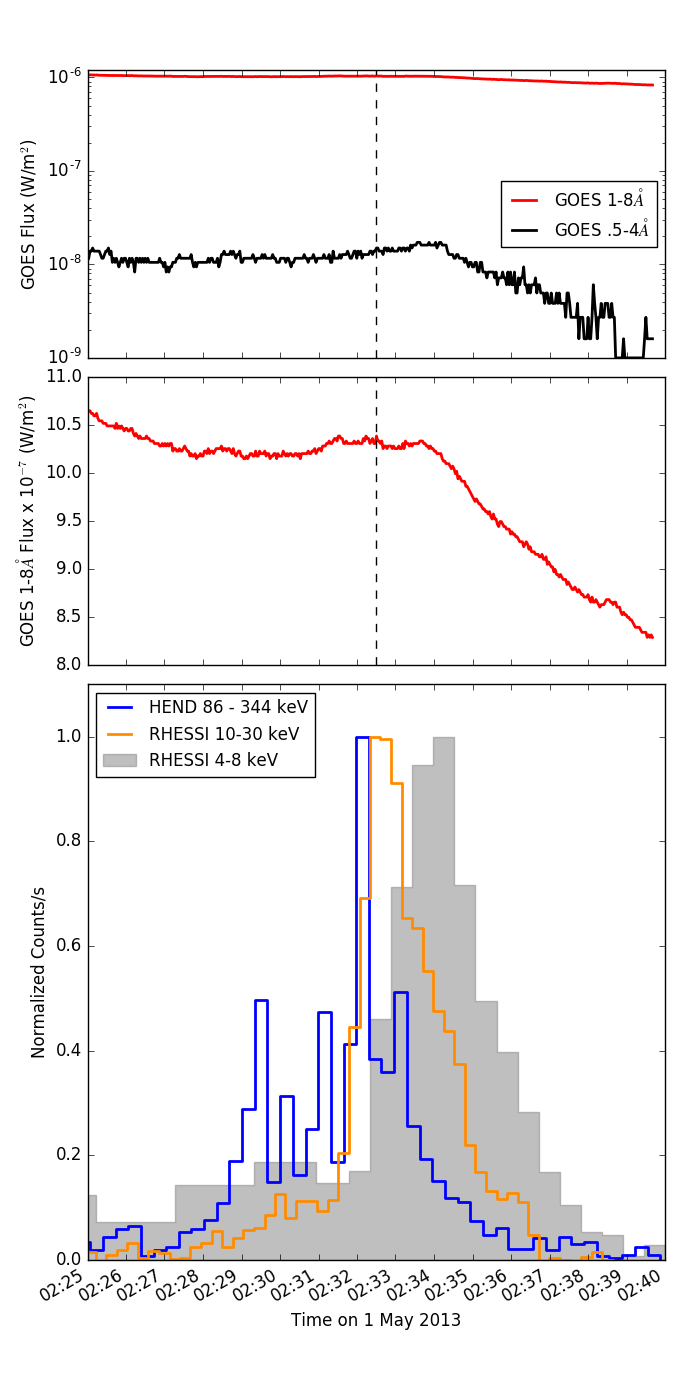}
\caption{Left panels: Lightcurves from instruments sensitive to hot plasma (top) and cooler plasma (bottom), for the same time period shown in the animated version of Figure \ref{fig:movie}. 
Flux was determined from an area 400" x 400" either centered on the flare (direct perspective of STEREO-B) or above the limb but in the same vertical extent (all occulted perspectives).
Each lightcurve was background-subtracted using pre-flare values and normalized relative to the flare peak.
The symbols indicate the imaging time cadence, which influences the relative timing of the peaks, especially in the bottom panel. The shading indicates the time period illustrated in further detail on the right.
Right panels: GOES, RHESSI, and HEND lightcurves. The GOES long channel, shown in linear scale (middle) shows an increase of flux corresponding to the RHESSI  emission, overlaid on a decaying slope. The direct view of HEND observed bursts before the main flare onset.}
\label{fig:lightcurves}
\end{figure*}

Figure \ref{fig:movie} shows views from various instruments near the peak of the flare (02:32:28UT in RHESSI 13-30 keV), and the accompanying animation shows the four-hour period surrounding the event for context. Lightcurves, shown in Figure \ref{fig:lightcurves} for selected instruments are shown over that same time period (left panels) or in detail when the HXR emission occurs (right panels).

EUV views from the occulted perspectives - AIA and STEREO-A - showed varying amounts of increase in the hour leading up to the flare. The direct view of STEREO-B showed loops brightening and changing over the active region as it evolved. This behaviour is consistent with the appearance of loops over the solar limb as seen in the movie, and was accompanied by an increase in soft X-ray flux as seen by GOES SXI. After the flare peaks, dimming was seen in many EUV wavelengths, including material seen in absorption by STEREO-B. Shortly after the flare peak, STEREO-A presented a beautiful view of an erupting filament in both its channels, with small, bright plasmoids running up the axis of the filament. This was less spectacularly seen by AIA's EUV channels, perhaps due to the higher occultation. Post-flare loops were slow to emerge, with the first candidates being seen at 03:00UT by STEREO-A. Likely the high occultation angle, combined with an unfavorable loop orientation, allowed only the tops of the highest cooling loops to be seen from the occulted perspective. 

In the X-ray regime, flux initially decreased as a previous flare on the solar disk decayed. However, after 2:20UT, RHESSI detected a second fainter source above the limb that would peak little over ten minutes later.
The RHESSI 10-30 keV peak occurred at 02:32:28UT and the 4-8 keV peak was slightly delayed at 02:34:15UT (see right panel, Figure \ref{fig:lightcurves}). The High Energy Neutron Detector (HEND), part of the Gamma Ray Spectrometer aboard Mars Odyssey, saw the main peak at 02:32:09UT in channels from 65-200 keV. It observed several bursty peaks starting five minutes before the main peak\footnote{This flare was not listed in the HEND flare catalog \citep{livshitsCatalogHardXray2017a} because the corresponding GOES flux increase was not enough to indicate a flare due to the occultation.}. GOES XRS did not register this emission as a flare. At the time of the RHESSI HXR peak, the amount of flux in the 1-8\AA{} channel corresponded to a C1 background with a B1 enhancement.

Because the majority of the event was occulted as seen from Earth, we used the STEREO-B EUV images to better estimate the flare magnitude. Following the method of \cite{chertokSimpleWayEstimate2015}, the 286" length of the saturated portion of the peak STEREO-B image at 02:30:57UT gave a flare class of M7. 
The method of \cite{nittaSoftXrayFluxes2013}, which compares the full-disk EUV flux before and during the flare, gave a GOES class of M3. 
Both these empirically derived formulae have large uncertainties, with Chertok claiming an accuracy of less than a factor of 2 and Nitta within 0.5-1.5 times the actual SXR flux value. Because the STEREO peak image occurred almost three minutes before the peak SXR flux, it is safe to assume that the flare was at least a moderate M-class but  not as large as X-class.

RHESSI imaging, which will be discussed in more detail in sections \ref{sec:th-img} and \ref{sec:nt-img}, showed both thermal and non-thermal sources. 
The thermal source was visible for 5.5 minutes, outlasting the non-thermal source which could only be imaged over three minutes.
The non-thermal source appeared both higher in the corona and earlier than any emission shown by the NoRH 17 and 34 GHz images, which corresponded well with the  AIA 304\AA{} images, showing a large expanding bubble which lasted long after the X-ray sources faded. 
Both X-ray sources were clearly located behind the CME front. The CME itself, as seen by AIA, had a multi-thermal structure whose 3D nature was unclear due to projection effects. The hotter FeXVIII image showed most material concentrated in a single loop towards the north, while cooler plasma was distributed much further south and higher above the limb. 

\section{Spectral Analysis}\label{sec:spectra}

\begin{figure}
\includegraphics[width=.5\textwidth]{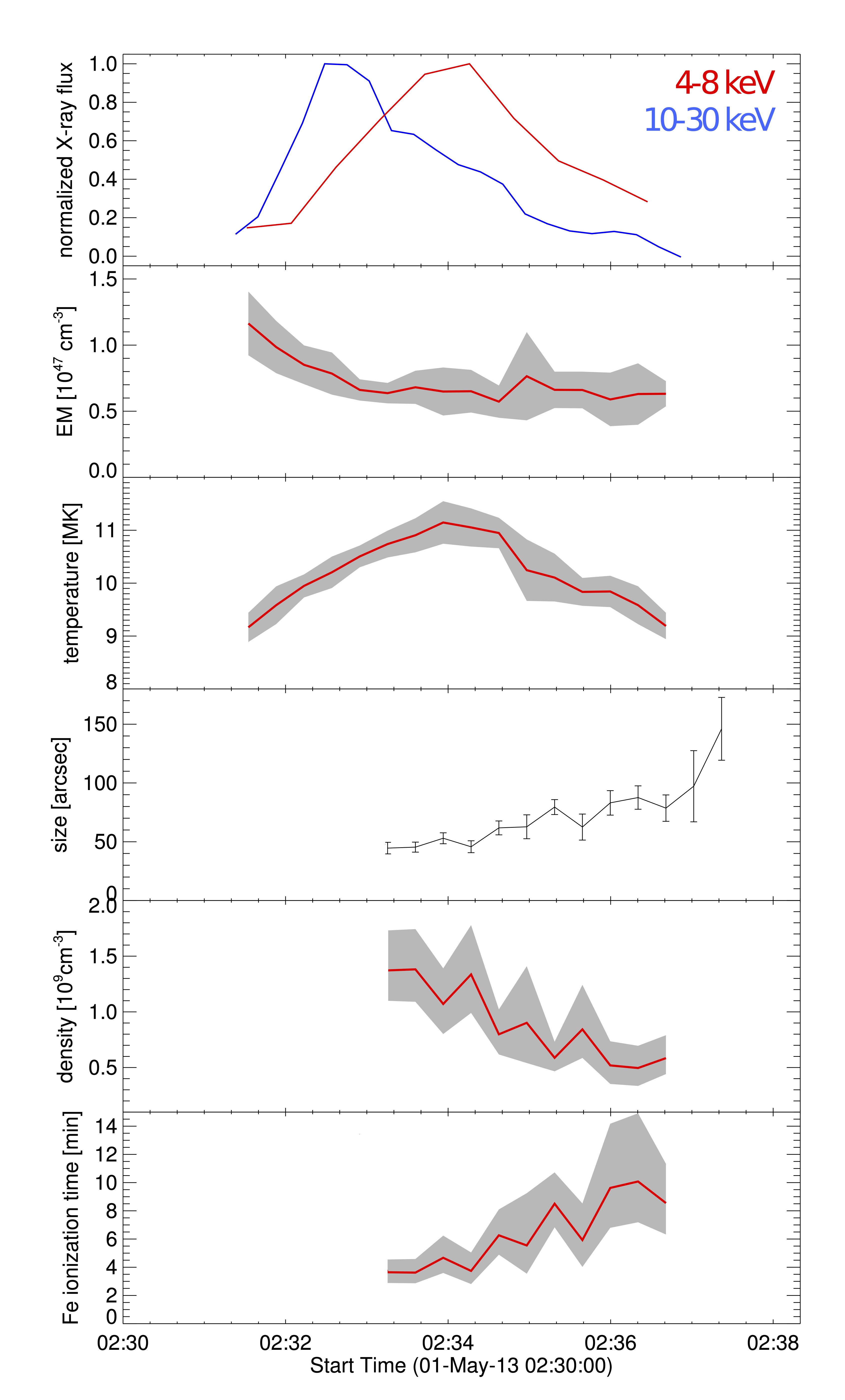}
\caption{Properties derived from the RHESSI SXR source (red curve, top plot). The emission measure stayed surprisingly constant, although the temperature followed the expected behaviour. The last three panels, which depended on imaging, are shown when the flare fluxes clearly exceed the flux of the on-disk source. The compact thermal source almost doubled in size as it rose above the flaring region and the solar limb. The density, calculated using a filling factor of unity, follows the temperature profile. Long iron ionization times might explain why AIA did not observe the thermal source.}
\label{fig:rhessi_props}
\end{figure}

\begin{figure}
\includegraphics[width=.5\textwidth]{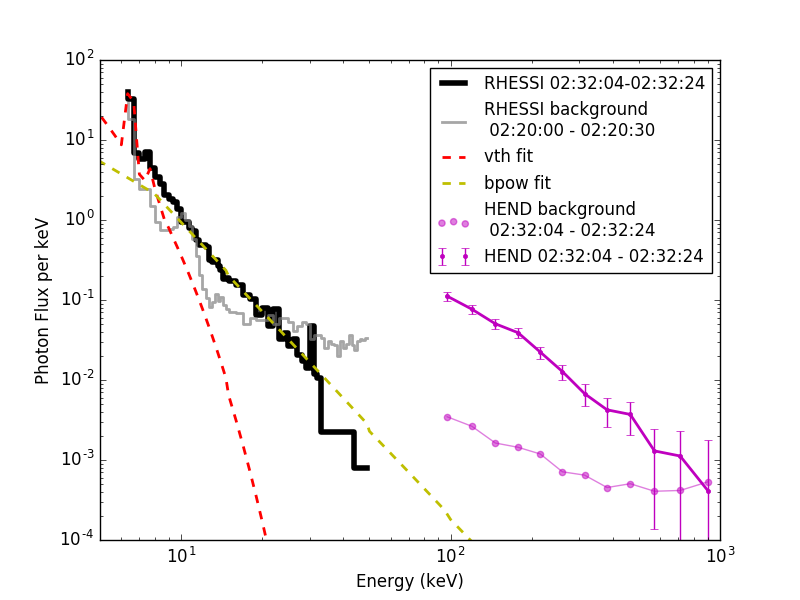}
\caption{RHESSI and HEND photon spectra. The RHESSI photon spectrum was fit with both a single thermal component plus a broken power-law.  
}
\label{fig:rhessi}
\end{figure}


\subsection{RHESSI thermal spectra}\label{sec:thermalspec}

Spectral analysis of the RHESSI thermal emission was complicated by the simultaneous occurrence of a decaying microflare located on disk. This made it difficult to fit the total flare spectrum integrated over the entire Sun; cross-talk at low energies resulted in a poor fit of a single thermal component. 
In order to cleanly isolate the two widely separated sources, we applied standard RHESSI imaging spectroscopy (\citet{kruckerRelativeTimingSpectra2002}). Using the coarsest subcollimators to distinguish coronal from on-disk emission 
enabled us to more accurately determine the properties of the thermal component. 

Fitting a single temperature model gave the temperature and emission measure (EM) evolution shown in Figure \ref{fig:rhessi_props}, with moderately high temperatures peaking around 11 MK. The EM time evolution was found to be very different from that of a standard flare. Instead of a fast decay following the flare peak, the emission measure, shown here in linear scale, had unusually small changes. This highlighted the fact that we were not observing the expected flare loops, but emission from well above the actual flare site. This was also influenced by the motion of the thermal source and the associated change in the degree of occultation. In any case, it is not clear how to explain the nearly constant EM.

We estimated the volume of X-ray emitting thermal plasma by forward fitting a circular source  to the RHESSI visibilities of all detectors (e.g. \citet{dennisHARDXRAYFLARE2009}). 
This revealed an extended source with FWHM sizes above $\sim$60" for all times, and a clear trend of an increasing size after 02:33UT.  Assuming a spherical symmetry and filling factor of unity, the derived density of the hot ($\sim$11 MK) plasma was around 10$^{9}$ cm$^{-3}$, a plausible density for this CME core in the high corona. The filling factor is an unknown; RHESSI observations, with a limited number of measured visibilities, cannot distinguish between an extended source and a composite of many subsources. Introducing a filling factor could drastically increase the density; however, as we will later discuss, a filling factor of unity agrees well with the data for this particular event.

\subsection{RHESSI \& HEND Hard X-ray Spectra}\label{sec:hxr-spec}

This flare was one of very few where spectroscopic observations of both the chromospheric footpoints and the coronal source were available. 
Measurements of the footpoint emission came from the High Energy Neutron Detector (HEND), part of Mars Odyssey's Gamma Ray Spectrometer (GRS) instrument suite \citep{boyntonMarsOdysseyGammaRay2004}, which observed energetic particles from $\approx$32-2000 keV. 
At the time of observation, the spacecraft structure did not shade the detector, simplifying interpretation of the detected flare X-ray photons.
The peak in the observed count spectrum was strongest in the 86-344 keV range. 
Even though this was a M-class flare, there were more than enough counts to clearly both distinguish solar emission from the background up to 400 keV, and render the errors negligible. 

Figure \ref{fig:lightcurves} shows that the RHESSI and HEND emission peaked almost simultaneously for the high-energy channels. 
During calibration, the HEND data was adjusted for the light travel time between Mars and Earth orbits. The HEND $>$ 90 keV peak, beginning at 02:32:09UT with a time binning of $\approx$20 seconds, might well have overlapped the RHESSI 10-30 keV peak in the 4-second rotation from 02:32:28-02:32:32UT. HEND showed multiple smaller peaks indicating high-energy bursts in the chromosphere starting at 02:28UT, before RHESSI registered the coronal source. 

An OSPEX fit of the photon spectrum integrated over the main non-thermal peak (02:32:00-02:33:13UT) to a power law with a low-energy cutoff gave slopes around $\gamma_{cor}$=3.3$\pm$0.38, where the uncertainties were derived from the standard deviations between the fit results of the individual RHESSI detectors. 
The flare-integrated spectra showed that non-thermal emission was the main component above 13 keV and extended down to at least 10 keV, below which thermal emission dominated. 
A single power-law fit to the HEND spectrum, excepting the two highest energy bins which had significant uncertainty, gave a spectral index of $\gamma_{chrom}$=2.47$\pm$0.26. The difference between the spectral indices of the coronal source (RHESSI) and footpoints (HEND) is within the range observed by \citet{battagliaRelationsConcurrentHard2006a} 
for flares with distinct above-the-looptop coronal and footpoint X-ray sources. Extrapolating the RHESSI fit to 100 keV, where the observed signal is difficult to distinguish from the background, we see that the photon flux observed by HEND was $\approx 10^3$ more than that observed by RHESSI.

\section{Coronal Source evolution}

\begin{figure*}
\includegraphics[width=\linewidth]{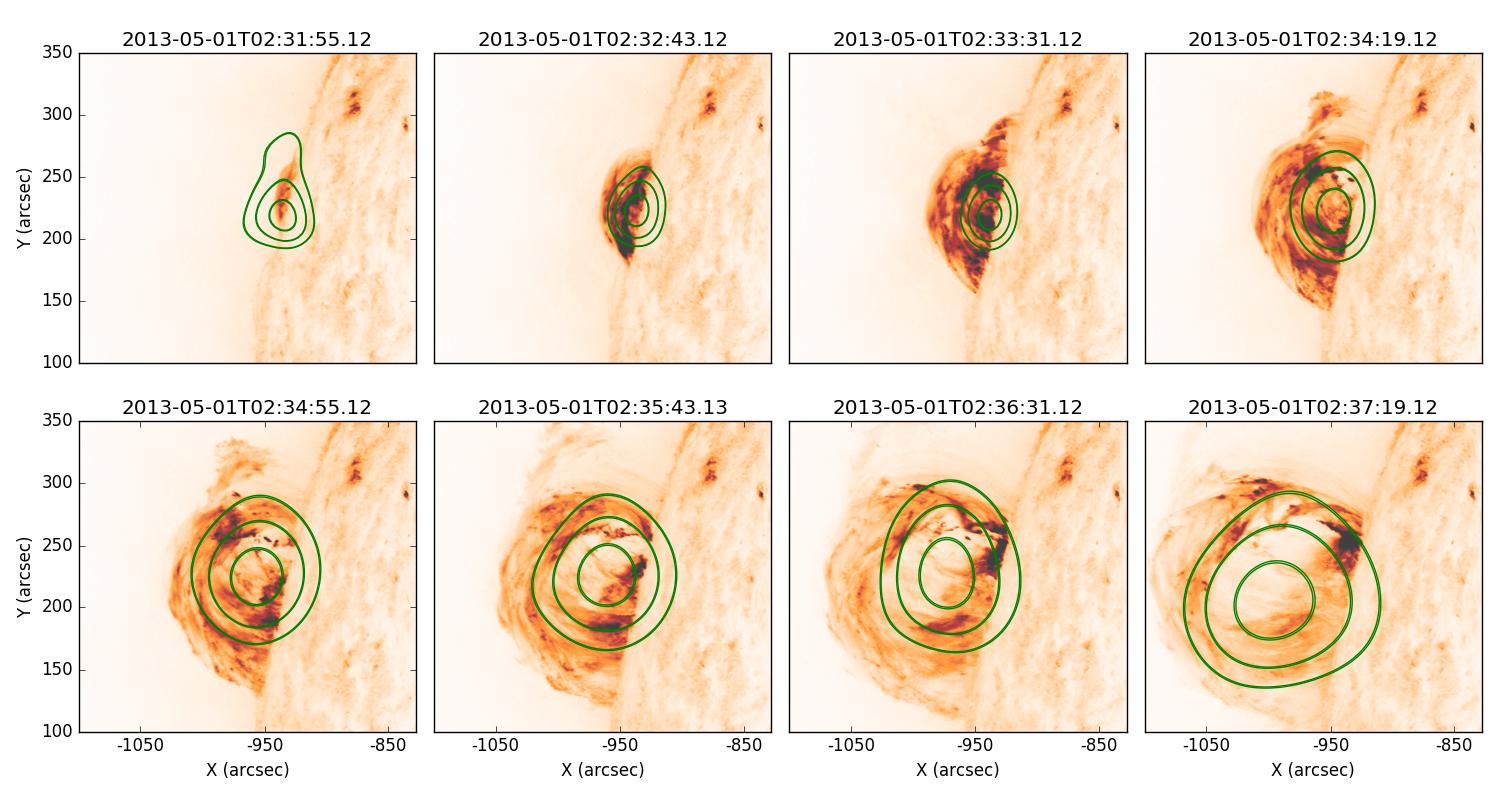}
\caption{Time evolution of the soft 4-8 keV X-ray source. Contour levels are 50, 70 and 90\%, as is common with RHESSI imaging. SXR images, made every 45 seconds starting at 02:32UT, show that the compact source expanded as it moved upward above the flare region. 
The background images are from AIA 304 \AA, corresponding to the timestamps.}
\label{fig:hxr_evolution}
\end{figure*}

\begin{figure*}
\includegraphics[width=.5\textwidth]{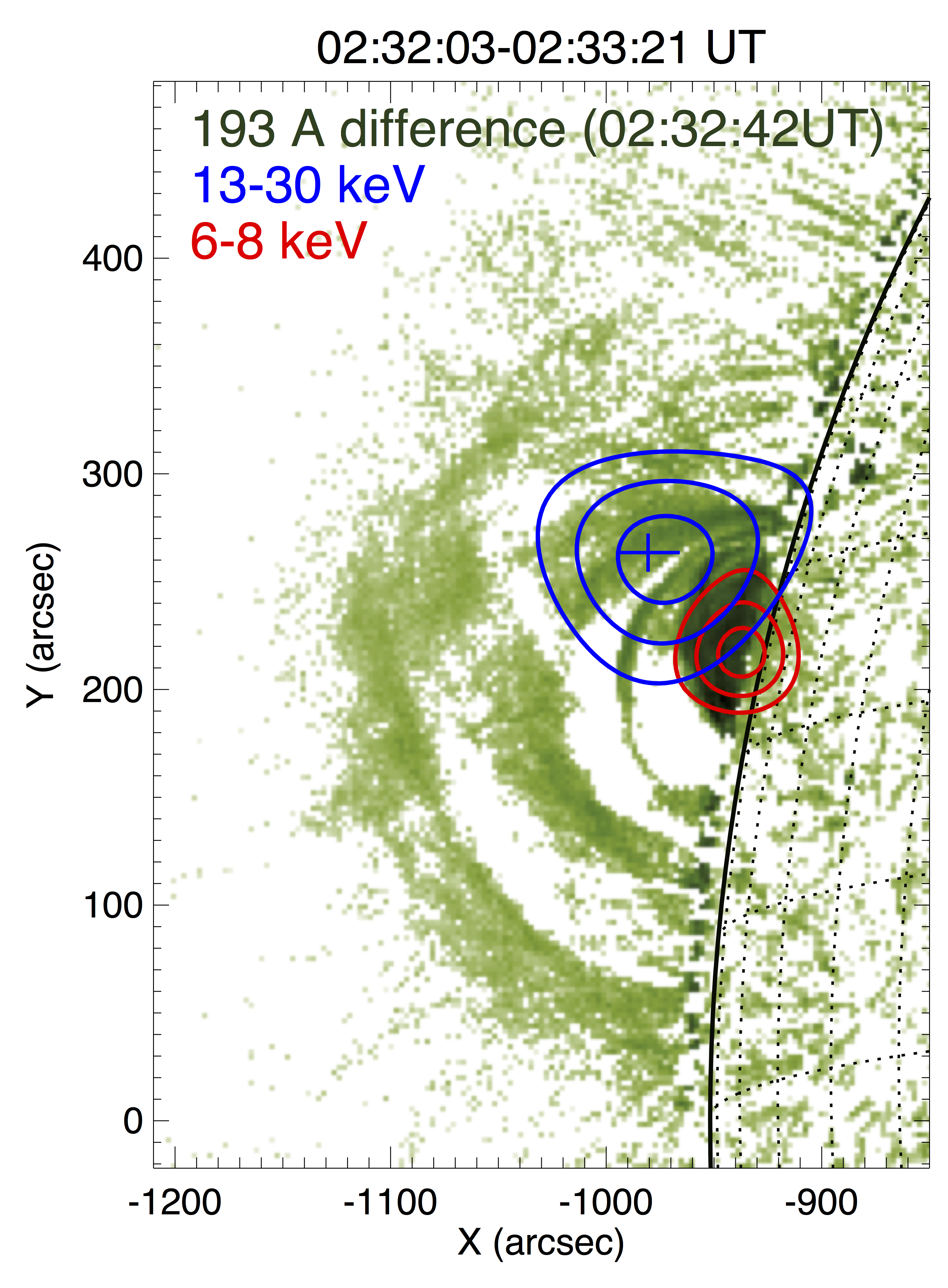}
\includegraphics[width=.5\textwidth]{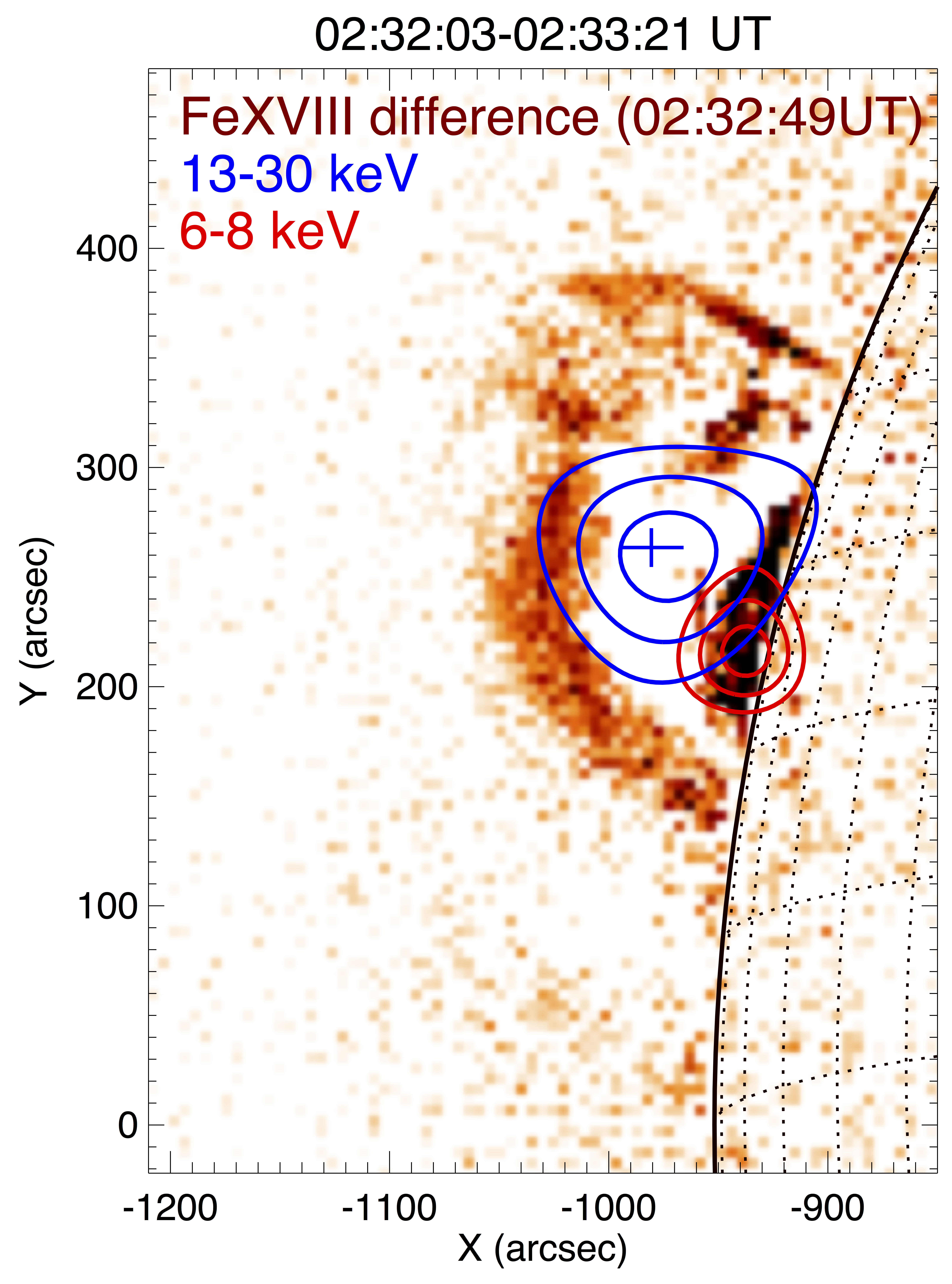}
\caption{The RHESSI HXR source is best summarized by this image. Low counts meant that a multi-step process was needed to ensure an accurate source size and position. During the flare peak, the extended HXR source lead the compact SXR source, both of which were high above the flare loop arcade but behind the CME front. Note that the CME front, likely due to projection effects and the 3-D structure of the event, appears to be located more to the north in the hotter FeXVIII image than for the cooler 193 image. }
\label{fig:rhessi_smooth}
\end{figure*}

\subsection{RHESSI imaging of the thermal emission}\label{sec:th-img}
We used RHESSI forward fitting to determine the size and position of the 4-8 keV X-ray source.
Not only did this allow us to separate the on-disk and coronal source, but also to determine the optimal parameters for imaging the thermal source with CLEAN. First, the source size and position  was calculated by forward fitting. Next, images for each subcollimator separately were reconstructed using CLEAN. 
If the source position agreed with the position calculated by the forward fit, that subcollimator was selected for use in making the final CLEAN images shown in Figures \ref{fig:hxr_evolution} and \ref{fig:rhessi_smooth}. 

The CLEAN images show the expansion and movement of the compact thermal source. The imaging, when overlaid on AIA 304 \AA{}, clearly showed that the thermal emission came from the core of the CME. This particular AIA filter sees cold plasma, which exhibited fine structure within the CME core. In this event, cold and hot plasma must have co-existed in the core, at least in projection.

A fit to the 4-8 keV centroid positions over time gave an average velocity of 150 km/s. This was much slower than the linear CME speed, measured by LASCO as 389 km/s. The source rose to a maximum height of 73 Mm above the limb, or 184 Mm (243") above the active region. 
This is one of the best RHESSI image sequences to date of a purely coronal source from within the core of a CME rising and expanding above the limb.

\subsection{RHESSI imaging of the non-thermal emission}\label{sec:nt-img}

From the RHESSI spectral analysis (section \ref{sec:hxr-spec}), we concluded that emissions above 13 keV were purely non-thermal. Above 30 keV, the background began to dominate the count spectrum. We therefore selected the energy range from 13-30 keV for imaging the non-thermal source. Relative to the thermal source, count statistics were low, with only about 1000 counts per detector if integrated over the entire non-thermal peak, so to summarize  we reconstructed a single image averaged over the duration of the non-thermal burst (02:32:03-02:33:21UT).
For reference we made an image of the thermal source in the further restricted 6-8 keV energy range integrated over the same time interval. 

Using a forward fit, we found that the thermal source came from an extended (61$\pm$10" FWHM) area. The non-thermal emission came from an even larger (110$\pm$30" FWHM) area. The center of mass location difference for the thermal and non-thermal source of 41$\pm$13" in the x- and 39$\pm$10" in y- direction, or a radial separation of 56.6$\pm$16".
This clearly established that the non-thermal source was above the thermal CME core. For the summary image shown in Figure \ref{fig:rhessi_smooth}, we then used the CLEAN algorithm to make images without restricting the source shape, again allowing the source size derived from forward fitting to guide the subcollimator selection. We chose subcollimators 6 through 9 for the thermal source and subcollimators 7 through 9 for the non-thermal source. 

 Note that the limb in EUV is slightly higher ($\sim$10") than the X-ray limb (e.g. \citet{battagliaSolarXRayLimb2017a}), so the X-ray source appeared at slightly lower altitude. The non-thermal source was clearly outside the CME core and just ahead of it, but still much behind the CME front even with its multi-thermal structure. 

\subsection{Nature of the non-thermal coronal emission}\label{sec:nature-nt}

We estimated the instantaneous number of electrons that are required to produce the observed hard X-ray emission using equation 2.4 from \citet{linNonrelativisticSolarElectrons1974}. Such an estimate depends on parameters of the observed hard X-ray spectrum and the ambient density within the non-thermal source - that is to say, the thermal core distribution. While the absolute value and the slope of the power spectrum was well observed, the cutoff energy of the non-thermal spectrum and the ambient density are not well-constrained by the observations. Hence, the instantaneous number of non-thermal electrons and therefore the instantaneous number density can only be estimated for a range of parameters. 

\begin{figure}
\includegraphics[width=.5\textwidth]{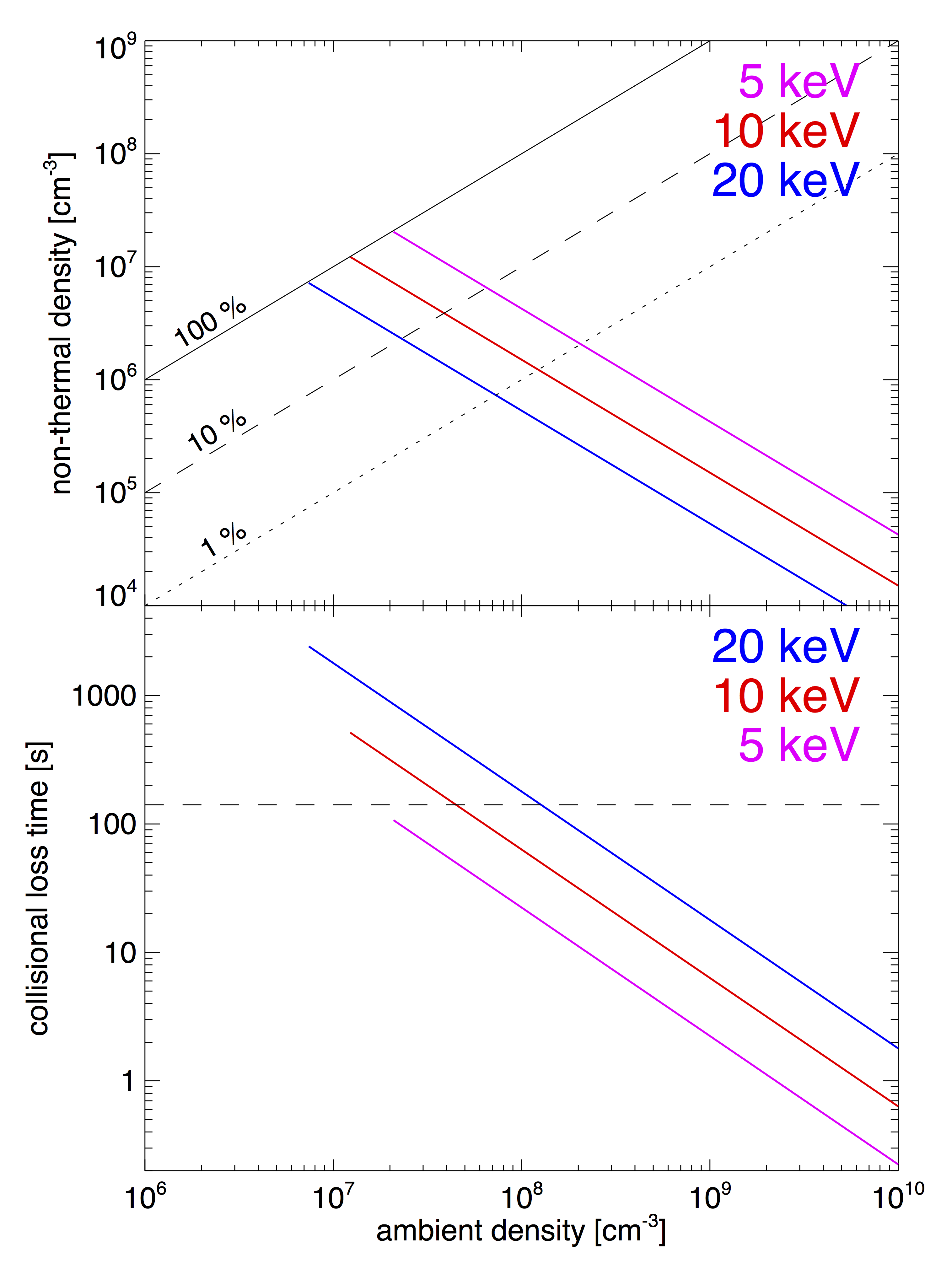}
\caption{Non-thermal electron density (top) and collisional loss timescales (bottom) calculated for bremsstrahlung emission at different energies.}
\label{fig:ambient_density}
\end{figure}

Figure \ref{fig:ambient_density} (top) gives the density of the instantaneous electrons as a function of ambient density for three different cutoff energies. The solid line represents the extreme case where the non-thermal density is equal to the ambient density. In this case, the derivation is no longer valid since collisions between non-thermal electrons should be considered as well.  Generally, the non-thermal component is thought of as a tail on the thermal core distribution that only represents a small fraction of the total particles in the distribution. The ambient density within the non-thermal source is not well constrained, but due to its higher altitude it should be below or possibly as high as the density of the hot core (~10$^9$ cm$^{-3}$, see Figure \ref{fig:rhessi_props}). 

An estimate of the ambient density within the non-thermal source can be approximated if we assume that non-thermal electrons are trapped within the source and the observed decay of the non-thermal emission is due to collisional losses only. In this case, the collisional stopping time (e.g. \citet{kruckerHardXRayEmissions2008}, eq 2) should be roughly equal to the observed exponential decay time of 141 seconds (13-30 keV). The bottom panel of Figure \ref{fig:ambient_density} shows the energy loss time as a function of density for different electron energies. This indicates that densities around ~10$^8$ cm$^{-3}$ give plausible energy loss times. 

An alternative theory is that the observed flux decay results from a rapidly decreasing ambient density as the bubble trapping the non-thermal electrons expands. In such a model, the thin target emission, which is proportional to the ambient density, decreases in time. Therefore the hard X-ray flux also decreases.  The collisional losses also decrease with time, meaning that the non-thermal electrons could potentially survive for a long time within the escaping bubble. For a constant, isotropic expansion velocity,  density decreases with time proportional to the third power. As the exponential fit mentioned previously is only observed over a short time interval, the decay can also be approximated with a power-law decay, at least for this event with limited counts. For isotropic expansion velocities of the order of 200 km/s, the observed decay can be roughly reproduced, assuming the injection stops at the peak time of the non-thermal emission and using the observed source size at that time of 110". As the CME is expanding with roughly double that speed (see Table \ref{tbl:events}), it is appears possible that the escaping bubble could move out at 200 km/s. For events with better statistics than for the May 1 event, the fit to the decay time could clearly distinguish between the two scenarios --- collisional decay vs expansion. For the October 27 flare \citep{kruckerSolarFlareHard2007} , which has 30 times higher count rates (see Table \ref{tbl:numbers}), an exponential decay is clearly preferred (Figure 1 and 3 of \citet{kruckerSolarFlareHard2007}). Nevertheless, as the bubble of non-thermal electrons is most likely expanding during the time scale of the collisional losses, the effect of decreasing ambient density should be considered in addition to collisional losses.

In summary, we get a consistent picture assuming that the non-thermal emission is produced by a trapped population of energetic electrons within a plasma with an ambient density of $\sim$10$^8$ cm$^{-3}$. For such densities, the fraction of the non-thermal population is reaching values on the order of a percent (c.f. Figure \ref{fig:ambient_density}, top). Considering that the non-thermal electrons at 20 keV have roughly 100 times more energy than an average electron in the ambient corona at 2 MK, the energy content of the non-thermal population could be similar to the ambient energy and therefore might play a significant role in the total evolution of the event. 

\subsection{Coronal source at other wavelengths}

\begin{figure*}
\includegraphics[width=\textwidth]{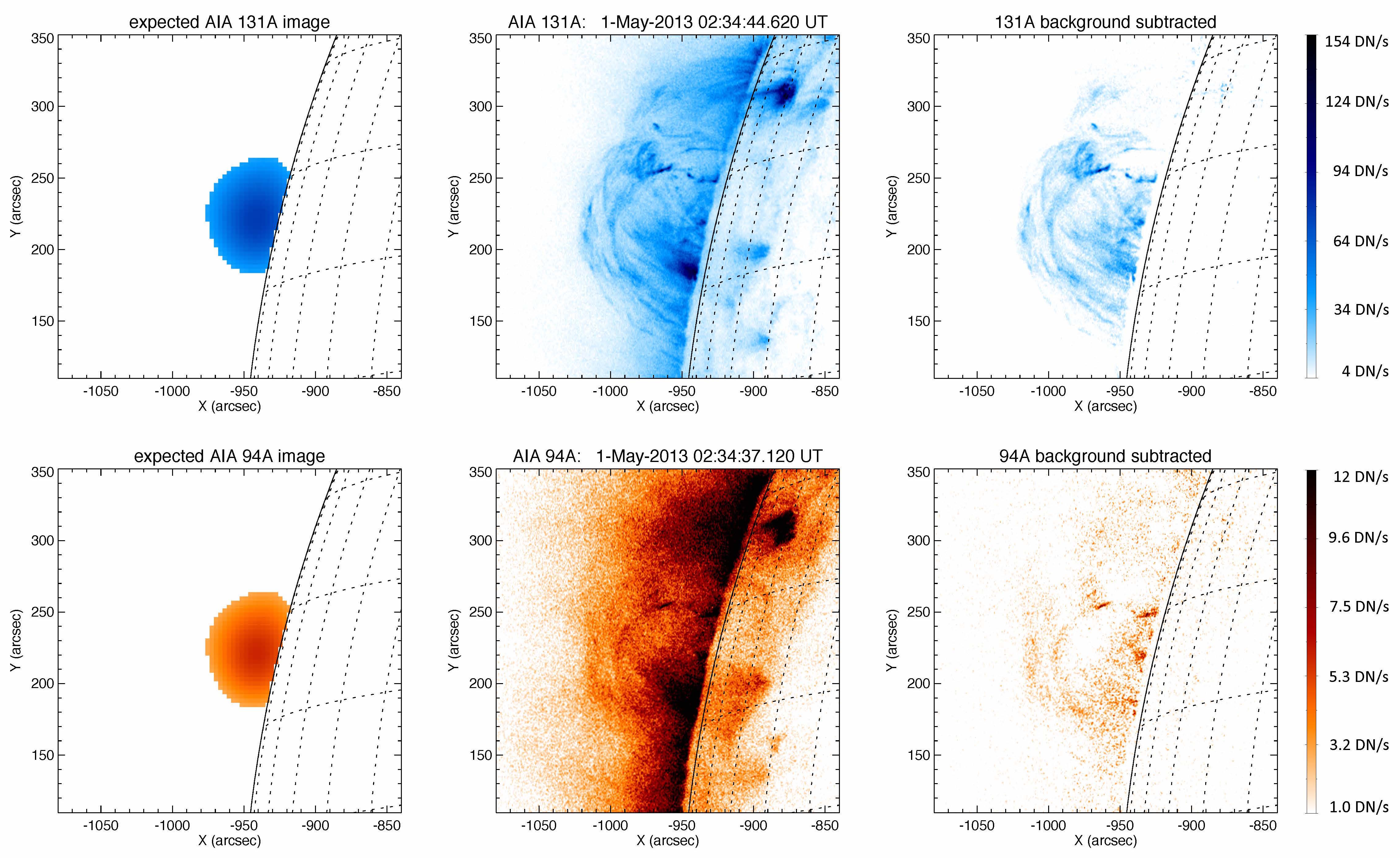}
\caption{Left: Brightness of expected sources in AIA 131 and 94\AA{}, given the size, location, and electron density of the thermal X-ray source. Middle: AIA observations. Right: Pre-flare background-subtracted observed emission.}
\label{fig:aia}
\end{figure*}

Although the pre-flare phase lightcurves showed flux increasing in AIA's hot channels, AIA images during the flare peak did not show a source in the corona corresponding to the thermal X-ray source. This is unlike the similar November 3 flare studied in \citet{glesenerObservationHeatingFlareaccelerated2013a}, which clearly showed the 11 MK CME core (the same temperature as derived for the CME core in this event). This is near the hot peak in AIA 131\AA's response function, due to FeXX and FeXXIII emissions. At this temperature ionization time scales for a low density plasma are not instantaneous. Ionization takes a few seconds or even minutes (e.g. \citet{bradshawCollisionalRadiativeProcesses2014}). Using the curves given in \citet{smithIonizationEquilibriumTimescales2010} for elements in constant electron temperature plasmas, we estimated the iron ionization time scales to be several minutes for this particular flare (see bottom panel of Figure \ref{fig:rhessi_props}). Therefore, the 131\AA{} signal associated with the RHESSI thermal emission is expected to have been delayed or even suppressed. 

We also searched for signs of the Fe line complex around 6.7 keV in the RHESSI data. Although the count spectra suggest that there might be a faint Fe line feature present, spectral fitting is inconclusive and does not give a quantitative result as it is unclear how to subtract the on-disk emission (see Section 3.1), making fitting the Fe line even more difficult than it already is for data taken late in the RHESSI mission lifetime when significant radiation damage to the detectors has been accumulated (these observations were taken 14 months past the last RHESSI anneal in February 2012).

Because the source was expanding while the density was decreasing, it is not straight forward to predict the actual flux that would have been observed in AIA 131\AA.  

We nevertheless made a simple calculation to determine the hypothetical EUV intensity, assuming instantaneous ionization. Given the size and temperature of the soft X-ray source, we should see 550K data numbers (DN) per second over the area of the source in 131\AA. In 94\AA{} this number is much less, at only 41K DN/s. With the area calculated from the X-ray imaging, both channels should show a source whose intensity is easily observable above the background, once the plasma has become sufficiently ionized. Figure \ref{fig:aia} shows what we could expect versus what was observed. The minimum ionization time of four minutes was on the order of the decay time of the RHESSI thermal emission. Other factors, such as the presence of flows or collisions with non-thermal particles might affect the time it takes for ionization equilibrium to be reached. 
However, because there are no signs of the plasma in AIA, this means that whatever the rate, the ionization simply could not catch up, regardless of the presence of non-equilibrium effects. 

We also calculated the expected 17 GHz gyrosynchrotron flux from the size, spectral index, and density of the non-thermal HXR source, using formula 2.16 from \citet{whiteRelationshipSolarRadio2011}. Only 1-10 sfu would be produced;
From the temperature and emission measure calculated from the thermal X-ray source, the potential 17 GHz emissions would be even less; using equation 15 from \citet{morgachevContributionThermalBremsstrahlung2014}, the contribution from the free-free emission would be only 0.035 sfu.
Imaging at 17 GHz using data from the Nobeyama Radioheliograph (NoRH) showed much stronger emissions. 
Ten-second cadence images, made at both 17 and 34 GHz (17 GHz is shown in the bottom-right panel of Figure \ref{fig:geom}), showed a bubble of cold material in thermal emission emerging over the limb and rapidly expanding. This cold material showed fluxes of up to 1000 sfu, bright enough to eclipse any emission from the same electron population producing the HXR source.

When the observed radio contours were overlaid on those from AIA's 50,000 K 304 \AA{} channel, the correlation immediately suggested that the majority of the emission seen in the radio was in fact due to thermal radiation. 
Coronal sources observed at GHz wavelengths have been interpreted as bubbles filled with nonthermal electrons \citep[c.f.][]{whiteRelationshipSolarRadio2011}; however this does not agree with the majority of the thermal emission observed in the radio for this event.  

\section{Discussion}

The earliest report of a purely coronal hard X-ray burst with a very hard spectrum reaching up to hundreds of keV was the famous event of March 30, 1969 \citep{frostEvidenceHardXRays1971}. Although not directly mentioned in that paper, the flare was occulted \citep{badilloSolarMicrowaveBurst1969}.
Using stereoscopic spectral observations of the February 16, 1984 flare, \citet{kaneStereoscopicObservationsSolar1992} 
were the first to determine that HXR sources must be extended, greater than 100” in size.
The first observations with imaging were provided by Yohkoh HXT \citep{hudsonHardXRadiationFast2001}, which confirmed that the coronal source was extended and furthermore, moving away from the Sun.
The best imaging so far was done by \citet{kruckerSolarFlareHard2007} using RHESSI, which clearly showed large sources that expanded while moving away from the Sun. Eventually the sources inflated to sizes so large that even the smallest Fourier component observable by RHESSI did not have a significant signal. 
\citet{glesenerObservationHeatingFlareaccelerated2013a} 
observed a related though barely-occulted event which demonstrated that HXR-producing electrons could heat the CME core to temperatures around 10 MK. Table 2 summarizes the key observables for previously published events.

 While it is not clear that all studies indeed describe the same type of event, they share several common characteristics:
\begin{itemize}
    \item The hard X-ray profiles are rather simple with a broad peak followed by an exponential decay. 
	\item The non-thermal part of the spectrum is generally hard and tends to further harden during the decay. 
	\item Sources are spatially extended and move away from the Sun while expanding.
	\item Sources are observed behind the front of the escaping CME.
\end{itemize}

\begin{table*}
\caption{Published properties of selected occulted flares}
\label{tbl:events}
\begin{tabular}{p{1cm}p{3cm}cp{1cm}cccccccc}
\hline
Date & Publication & Occultation & GOES & CME & 30 keV & $\tau$ (s) & $\gamma_{cor}$& $\gamma_{chro}$ & 30 keV& Size& Speed \\
&&&obs.&speed&photon&&&&flux flare&arcsec&(km/s)\\
&&&est.&(km/s)&flux&&&&&&\\
\hline
  Mar 30, 1969  &  \citet{frostEvidenceHardXRays1971}, \citet{badilloSolarMicrowaveBurst1969}  & 15\degree &&& $\sim$3 & $\sim$300 & $\sim$2 &&& \\ 
  Dec 14, 1971  &  \citet{hudsonPurelyCoronalHard1978}& 25\degree &&& 0.05& $\sim$600& 2.1 &&& \\ 
  Jul 22, 1972  & \citet{hudsonSecondstageAccelerationLimbocculted1982} & 20\degree &&& 0.1 & $\sim$400 & 2.5-1.8 &&& \\ 
  Feb 16, 1984  & \citet{kaneStereoscopicObservationsSolar1992} & 36\degree &B3 & & 0.4& $\sim$60 & 3.8-2.6 &3.3&150& $\>$ 140\degree & \\ 
  Jun 30, 1991  & \citet{vilmerHardXrayGammaray1991} & 2$\pm$12\degree & M5& &10 & $\sim$25 & 2.8-2 &&& \\ 
  Apr 18, 2001  & \citet{hudsonHardXRadiationFast2001}  & 27\degree &C2 \hspace{1cm} $>$ X1& 2400& &$\sim$30 & 4.3-3.4 &&&20"- 70"& $\sim$1000\\ 
  Oct 27, 2003  & \citet{kruckerSolarFlareHard2007} \citet{vybornovObservationPowerfulSolar2012}  & 40\degree &$\sim$B1 \hspace{.5mm} $>$ X1& 2300 & 0.1 & $\sim$135 &3.6-3.1 &2.3&80&200"&$\sim$750 \\ 
  Nov 3, 2012  &  \citet{glesenerObservationHeatingFlareaccelerated2013a} & 6\degree &C5 \hspace{1cm} X1& 240 & 0.3 & $\sim$30 &4.5&&&50"-100"& \\ 
  May 1, 2013  &  Lastufka et al. (2019) & 30\degree &B \hspace{1cm} M3-7& 400 &0.03& $\sim$140 & 3.3 &2.5&20&100"&$\sim$ 200 \\
  Sep 1, 2014  & \citet{carleyEstimationCoronalMass2017} \citet{grechnevRadioHardXRay2018} & 36\degree &$\sim$B1 X2&2000& 0.2 & $\sim$900 &2.06\footnote{This spectral index was derived from flux integrated over the entire flare \citep{ackermannFermiLATObservationsHighenergy2017b}, whereas $\gamma_{chro}$ and the 30 keV flux quoted are for the first X-ray peak seen by HEND}& 3.3 &220&$\>$200"&\\ 
  \hline
  \end{tabular}
\end{table*}

These facts support the current best explanation that these HXR sources are produced by bremsstrahlung emission from flare-accelerated electrons, which escape upward from the coronal acceleration region.
Flare-accelerated electrons cannot escape freely from the Sun because they are injected within the complex magnetic field structure of the CME core. Due to the low ambient density in the corona, these electrons only lose energy slowly; however, they can move around within the CME core during their livetime, resulting in a large source size.

The observed exponential decay is likely the result of the combined loss mechanisms after the injection stops. The largest contributor is probably Coulomb collisions, but electrons leaving the source volume can also result in a reduction of flux. Other loss mechanisms might be at play as well. It is also possible that acceleration might not stop completely after the event peak, such that continuing injection at a low rate prolongs the decay. 
Using estimates of the ambient density, the expected collisional loss times roughly agree with the observed time scales of the decay, indicating that collisions alone could be enough to explain the decaying time profile. 
The collisional losses are expected to heat the ambient plasma, creating a hot thermal source within the CME core \citep{glesenerObservationHeatingFlareaccelerated2013a}.

The observations of the May 1 flare discussed in this work fit well within this picture of large source sizes and exponential decay. 
What is different from the event in \citet{glesenerObservationHeatingFlareaccelerated2013a} is that the source of the non-thermal emission was spatially displaced from the hot CME core. 
Therefore, further evidence is required to conclude that the May 1 CME core was heated by non-thermal electrons.
If such heating did occur, it must have been from a HXR source co-spatial with the core that was no longer observable by the time the core became visible above the solar limb.
This is certainly possible; comparing with the event studied by \citet{glesenerObservationHeatingFlareaccelerated2013a}, the CME core would have been barely visible above the limb at the end time of the HXR burst (c.f. Figure 3, panel 6, from \citet{glesenerObservationHeatingFlareaccelerated2013a}) if it had had the same occultation height seen here for the May 1 event. 
However, this implies that the RHESSI non-thermal emission shown in Figure \ref{fig:rhessi_smooth} must have come from a different population of energetic electrons than that of the observed HXR source.
These could have been injected into a magnetic structure with a lower density, allowing the electrons to survive longer, or a second injection could have occurred at a later time. 
Nevertheless, we have no observational evidence for such a scenario so the prudent explanation is that that hot core in the May 1 flare was not heated by non-thermal particles, but rather a different mechanism.

For four events in Table 2, including the May 1, spectral observations of the main flare emission allow us to estimate the relative intensity of the coronal source with respect to the rest of the flare.
Spectral information for full-flare perspective was only available above 100 keV. High coronal events are best seen at lower energies around 30 keV, making an extrapolation necessary. 
Comparing the extrapolated fluxes at 30 keV, the total flare emission is 400 to 800 times stronger than the coronal emission. 
The flare spectra also tend to be harder in the chromosphere/low corona than in the high coronal source; hence, the difference in flux is expected to increase at higher energies with values between 1000 and 1500 at 100 keV. 

While these flux ratios are relevant for detection, the physical significance comes from the number of accelerated electrons within each source. To derive the number of electrons, we have to adopt a certain model. 
The total flare energy is usually derived by assuming that electrons lose all their energy in the chromosphere, with collisions being the main loss mechanism (e.g. cold thick target model, e.g. \citet{brownDeductionEnergySpectra1971}). This is a rather robust assumption for chromospheric sources, but it depends on low-energy cutoff that is poorly constrained. 
We summarize the energy input to the chromosphere by electrons using 20 keV as a reference in Table 3. While the coronal high-energy source is also likely produced by bremsstrahlung (for a discussion on the possible contribution of inverse Compton radiation see \citet{chenROLEINVERSECOMPTON2012}), the low ambient density within the source makes the classic thick target assumption inappropriate. Therefore the number of electrons needed to produce the hard X-ray spectrum at any time can be estimated from the instantaneous number of electrons by assuming a thin target model.

However, the instantaneous number $N_e$ depends not only on the observed photon spectrum (as does the thick target approximation), but also on the ambient density, to which it is inversely proportional. 
Using an ambient density of 10$^8$ cm$^{-3}$ for all events, Table 3 gives the instantaneous number of electrons at peak time, again with 20 keV as the reference energy.
$N_e$ at peak time is the maximum number of accelerated electrons that radiates at any time. 
It can therefore be used as a lower limit of the total electrons in the high-altitude source. The actual number could be larger, as a lower-magnitude injection of electrons could continue after the peak time. Particle-acceleration occurring in flares well after the peak is a well-established behavior. Additionally, a larger number of electrons might have been injected before the non-thermal source became visible above the limb as seen from Earth. Considering all these uncertainties, and that the ambient density is not well constrained, the estimate of the number of electrons is uncertain by a factor of a few at least. Nevertheless, it is our current best estimate that compared to the number of >20 keV electrons in the main flare peak, the number of electrons in the high coronal source are below the percent range.

\begin{table}
\label{tbl:numbers}
\centering
\caption{Estimates of the number of electrons above 20 keV assuming thick target emissions from the footpoints and thin target for the extended coronal source. }
\begin{tabular}{lcc}
\hline
Date& Thick target $N_e$ & Thin target $N_e$\\
& footpoints & corona\\
&$>$20 keV &$>$20 keV at peak\\
\hline
  Feb 16, 1984  & 2.5\rm{x}10$^{39}$ &2.5\rm{x}10$^{36}$ \\
  Oct 27, 2003  & 3.5\rm{x}10$^{38}$ &6.7\rm{x}10$^{35}$  \\
  May 1, 2013  & 9.4\rm{x}10$^{37}$ & 2.0\rm{x}10$^{35}$ \\
  Sept 1, 2014  & 2.7\rm{x}10$^{39}$ &7.7\rm{x}10$^{35}$  \\\hline 
\end{tabular}
\end{table}

\section{Conclusion}
The moderately sized flare and CME of May 1, 2013 was uniquely well-situated for high-energy observations associated with CMEs.  
Due to the event's position 30\degree{} behind the solar limb, RHESSI saw coronal emission with no contamination from the footpoints or flare loop arcade. 
From the opposite side of the Sun, HEND viewed the full, un-occulted flare. Radio and EUV instruments were available to provide context and constrain the interpretation of the event.

Analysis of the thermal X-ray emissions found that plasma with temperatures up to 11 MK was ejected. The hot source was located behind and rose slower than the CME front, indicating that it was likely the result of hot plasma trapped in the complex magnetic fields of the CME core. RHESSI imaging of this large thermal X-ray source expanding as it rose above the flare site may be the best such observation to date. The long iron ionization time scales for such a high altitude source at low density made the hot core of the CME undetectable in EUV, unlike in \citet{glesenerObservationHeatingFlareaccelerated2013a}.

Imaging also showed a clear short-lived non-thermal source, which was very extended with a FWHM of 110", located 185" in projection above the flaring region. This was both higher up than the thermal CME core, and still behind the CME front. It must have originated from a more tenuous part of the CME core where non-thermal electrons survived long enough to become visible from Earth.
Because of their location above the CME core, the non-thermal particles in this source could not have been responsible for heating the core itself. It is possible that a different population of non-thermal electrons heated the core to its 11 MK temperature, but these were not observed due to the occultation. Another possibility is that a different heating mechanism entirely was responsible. Either way, we were unable to say for certain how the CME core was heated. 

Assuming a thick-target model for the HEND spectrum and a thin-target model for that of RHESSI, we deduced that the number of coronal vs chromospheric non-thermal electrons to be one in five hundred. 
Although some bursts were observed in the flaring region before the main peak, both RHESSI and HEND observed the main electron acceleration at almost the same time. HEND, whose count spectrum was dominated by the chromospheric footpoints, saw a thousand times more photon flux than RHESSI. 
Both this flux ratio and the spectral index difference agrees with earlier studies by \citet{kaneStereoscopicObservationsSolar1992,vybornovObservationPowerfulSolar2012}; however, the flares examined in these works were extremely large. With X-ray imaging spectrometers, it is understandable that only the largest such events are identified, since larger flares produce more high-energy counts which improves the possibility and quality of imaging.
The presence of similar properties in the M-class flare of May 1 invites us to consider whether such a flux ratio and spectral index difference are common to all flares regardless of magnitude. Furthermore, the presence of coronal X-ray sources very high above the site of an average-sized flare supports the idea that these should be present in most flares, and could be revealed given sufficient observations from various angles, as will be provided by Solar Orbiter.
 
 \section*{Acknowledgements}
 \lasco

The work was supported by Swiss National Science Foundation (200021-163377) and through NASA contract NAS 5-98033 for RHESSI. The authors would like to thank the anonymous reviewer, whose comments improved the paper.

\bibliographystyle{plainnat}
\bibliography{May1.bib}

\end{document}